\documentclass[a4paper]{VisionStyle}
\usepackage{epsfig}

\begin{document}

\title{High Resolution X-Ray Spectroscopy of 14 Cooling-Flow Clusters of 
Galaxies Using the Reflection Grating Spectrometers
on XMM-Newton}

\author{J. R. Peterson\inst{1}, C. Ferrigno\inst{2}, J. S. 
Kaastra\inst{2}, F. B. S. Paerels\inst{1}, S. M. Kahn\inst{1}, J. G. 
Jernigan\inst{3}, J.~A.~M.~Bleeker\inst{2}, T. Tamura\inst{2} }

\institute{
Columbia University, Department of Physics, 538 W 120th St, NY, NY 
10027, USA
\and
SRON National Institute for Space Research, Sorbonnelaan 2, 3584 CA Utrecht, The 
Netherlands
\and
Space Sciences Laboratory, University of California, Berkeley, CA 94720, 
USA
  }

\maketitle

\begin{abstract}
We present high resolution X-ray spectra of 14 cooling-flow X-ray 
clusters and
groups obtained with the Reflection Grating Spectrometers (RGS) on 
XMM-Newton.
The spectra exhibit line emission from a number of Fe L charge states as 
well
as O VIII, Mg XII, Ne X, Si XIV \& XIII, N VII, and C VI.  All spectra 
show a
deficit of soft X-ray lines predicted from the isobaric multi-phase 
spectral
model as compared with morphological mass deposition rates from
spatially-resolved spectroscopy with the European Photon Imaging Cameras
(EPIC).  We present some weak detections of plasma several times cooler 
than
the ambient cluster temperatures.  The results further suggest that 
either
morphological mass deposition rates systematically overestimate the 
actual
cooling rate or the emission measure of cooling-flows has a much steeper 
distribution than that
predicted by a simple isobaric multi-phase model.  We briefly discuss 
some modifications to the cooling-flow
process.
\keywords{XMM-Newton -- RGS -- X-ray Spectroscopy -- Cooling-Flow -- 
Clusters of Galaxies}
\end{abstract}

\section{Introduction}

It has long been recognized that the cores of clusters of galaxies have 
sufficient X-ray luminosity to cool 10 to 1000 solar masses of keV 
plasma every year  (e.g. \cite{jpeterson-B3:fabian2}, \cite{jpeterson-B3:cowie}, \cite{jpeterson-B3:fabian1}).  The
details of the cooling process are still debated, however, in most 
models parcels of cooling plasma collect at the center
of the cluster, forming what is referred to as a cooling-flow.  Direct 
evidence for cooler gas in the cores of clusters is indeed
well-established (\cite{jpeterson-B3:canizares1}, \cite{jpeterson-B3:canizares2}, \cite{jpeterson-B3:mushotzky}).

If the gas were to cool homogeneously, the density profile of the 
cluster core would be relatively steep.  That is inconsistent with imaging observations (\cite{jpeterson-B3:johnstone}), which 
has led to the conclusion that the cooling gas must condense locally in
smaller clouds distributed over a large volume (tens of kpc), i.e. in a 
multi-phase medium.  However, even ignoring the details of the
resulting spatial distribution, simple thermodynamic arguments show that 
the integrated X-ray spectrum of such a cooling flow can be
robustly predicted.  If the blobs of gas cool in thermal isolation at 
constant pressure, and the dominant energy loss mechanism is via X-ray 
radiation, then the luminosity radiated per unit temperature interval 
must be proportional to the mass deposition rate, ${\rm \dot{M}}$:

\begin{equation}
\frac{dL}{dT} = \frac{5}{2} \frac{\dot{M} k}{\mu m_p}
\end{equation}

\noindent
where $k$ is Boltzmann's constant, and $\mu m_p$ is the mean molecular 
weight per particle.  The only free parameter is  ${\rm \dot{M}}$, which 
can be estimated from an X-ray image of the cluster core.  The resulting 
spectrum can be calculated using a collisional equilibrium spectral 
synthesis model with an assumed set of elemental abundances, and 
normalizing the contribution in each temperature interval as given in 
Equation (1).

Data acquired by the Reflection Grating Spectrometer on XMM-Newton have 
enabled this robust spectral prediction to be quantitatively tested for 
the first time.  Surprisingly, the observed spectra reveal a remarkable 
systematic deficit of emission at low temperature, as compared to the 
multi-phase model (\cite{jpeterson-B3:peterson1}, \cite{jpeterson-B3:tamura}, \cite{jpeterson-B3:kaastra}, \cite{jpeterson-B3:xu}).  
Emission is observed at the expected levels for temperatures just below 
the ambient temperature of the cluster, $T_a$ (down to roughly $\frac{1}{2}
T_a$), but not for significantly lower temperatures.  This result has been confirmed with medium resolution spatially-resolved 
spectroscopy using XMM-Newton EPIC and Chandra observations where spectral fits have yielded significantly smaller ${\rm \dot{M}}$'s than expected
(\cite{jpeterson-B3:david}, \cite{jpeterson-B3:boehringer}, 
\cite{jpeterson-B3:molendi}, \cite{jpeterson-B3:schmidt}, 
\cite{jpeterson-B3:ettori}).  The interpretation of this effect is still 
unclear, although a number of possible physical mechanisms for 
suppressing the low temperature X-ray emission have been suggested (e.g. 
\cite{jpeterson-B3:peterson1}, \cite{jpeterson-B3:fabian3}).  In this 
presentation, we review the results of RGS observations for 14 separate 
cooling-flow clusters, sampling a wide range
in mass deposition rate.  A companion analysis of the EPIC data for the 
identical sample of clusters is presented in \cite*{jpeterson-B3:kaastra2}.

\section*{Diffuse X-ray Spectroscopy with the RGS}

We briefly comment on the spectral capabilities of the RGS, and the 
analysis of RGS data for extended sources.  Since, the RGS is a slitless 
spectrometer (\cite{jpeterson-B3:denherder}), its spectral resolution is 
degraded for extended sources.  Nevertheless, in contrast to the
transmission grating instruments on Chandra, its dispersion is 
sufficiently high that it currently provides the highest spectral 
resolution for the soft X-ray band for sources of arcminute size.  The 
spectral resolution for an extended source larger than 10 arcseconds is 
given by

\begin{equation}
  \Delta \lambda \approx 0.1 { \rm \AA} \left({\rm 
source~size~in~arcminutes}\right) / \left( {\rm spectral~order} \right)
\end{equation}

At this resolution, one can still unambiguously resolve the most 
prominent resonance line contributions from all charge states of the 
most abundant elements.  Of particular importance is the the Fe L 
complex, which can be resolved into its emission line contributions from 
Fe XXIV through Fe XVII (Li-like to Ne-like).  The RGS resolution is several times higher than a CCD instrument in the Fe L region for a typical source size.

The combined effective area of the two RGS instruments is $280~{\rm 
cm}^2$.  The field of view is 5 arcminutes in the cross-dispersion 
direction and 1 degree in the dispersion direction.  This is 
sufficient to capture the entire cooling-flow region and produce a 
well-resolved spectrum with a 50 ks exposure for a typical cluster.  The 
instrument response for an extended source is the convolution of the 
spatial distribution of each emission line and the off-axis response for 
a given data selection region.  To properly account for these effects, 
we use a novel Monte Carlo code for all spectral fitting analyses 
(\cite*{jpeterson-B3:peterson3}).  A full spectral-spatial model of the 
source is used to set limits on the cooling-flow model.  An example of 
the RGS data and corresponding EPIC-MOS data is shown in Figures 1 and 
2.  Figure 3 shows a Monte Carlo simulation for the same data set.

\section*{The Soft X-ray Cooling-Flow Problem}

The failure of the isobaric multiphase cooling flow model to adequately 
describe the RGS data is illustrated for one of the most massive cooling 
flow sources, Abell 1835, in Figure 4.  The predicted spectrum (at the 
resolution of the RGS for this source) is shown in the top panel.  The 
model predicts detectable emission lines from the complete range of Fe L 
charge states, as well as from He-like and H-like charge states of N, O, 
Ne and Mg.  The comparison of this model with the observed RGS spectrum 
of Abell 1835 is shown in the second panel.  As can be seen, the fit is 
reasonable for the higher ionization Fe L lines (Fe XXIV and Fe XXIII), 
but the model vastly overpredicts the emission line intensities for the 
lower charge states, especially Fe XVII.  

The final panel shows an empirical parameterization, where the isobaric 
model temperature distribution has been arbitrarily truncated at 
temperatures below $kT = 2.7~{\rm keV}$.  This model has no clear physical 
motivation, but it does provide a good fit to the data.  Note that it 
reproduces the emission from Fe XXIV and Fe XXIII by the two line blends 
at 10.6 and 11.2 ${\rm \AA}$ roughly at the predicted value of ${\rm 
\dot{M}}$.

Previous, lower spectral resolution X-ray observations of cooling flow 
clusters had provided hints of this problem - a deviation from simple 
model predictions at the lowest energies.  However, the discrepancies to 
model fits in those cases were generally interpreted as evidence for 
absorption by cool clouds embedded in the cooling flow 
(\cite{jpeterson-B3:white}).  For the RGS data, we can measure the 
intensities of the emission lines individually.  It is clear from the 
raw data that absorption cannot explain this effect - the low energy 
lines are too weak, while the lower energy continuum is predicted 
correctly.  Further, the RGS data make it clear that the failure of the 
model cannot be ascribed to the assumed elemental abundances.  In 
particular, we see a deficit of low temperature emission lines in the Fe 
L spectrum alone.

There is, however, clear detections of large quantities of plasma at temperatures above $\frac{1}{2} T_a$ at roughly the predicted quantity.  This is very difficult to reconcile with any theoretical model.  This problem is essentially distinct from the classic 
cooling-flow problem of the failure to detect the final products of 
cooling-flows, molecular clouds and stars.  In fact, recent detections 
of CO emission by \cite*{jpeterson-B3:edge} from massive cooling-flows 
now suggest that the end products of cooling may indeed be present.  
However, some unforeseen process must be suppressing the cooling 
radiation in the soft X-ray band.

\section*{Other Clusters in the Sample}

Similar results for the full sample of 14 clusters are presented in 
\cite*{jpeterson-B3:peterson2}.
Generally, hot (several keV) clusters exhibit spectra similar to Abell 
1835 where no emission lines are detected below 3 keV.  Intermediate 
temperature clusters ($kT_a \approx 3$ keV) exhibit spectra similar to 
that  shown in Figure 5 for the cluster 2A0335+096.  Prominent emission 
lines from Fe XXII-Fe XXIV are very apparent between 10 and 12 ${\rm 
\AA}$.  Weaker emission lines are also detected from the lower charge 
states, however they become increasingly more discrepant with the 
predictions as the temperature at which they are produced drops to lower 
and lower values.  Since emission from essentially all Fe L ions is 
detected for these lower temperature clusters, the spectra clearly 
demonstrate the need for a continuous distribution of temperatures - but 
{\it not} the unique distribution predicted by the isobaric cooling flow 
models.

A spectrum from a 1 keV cool group of galaxies, NGC 533 is shown in 
Figure 6.  Here again, there is evidence for significant cooling below 
ambient temperatures.  The ratio of Fe XVII to Fe XVIII, however, is 
roughly 1, while the prediction for the standard isobaric model is 
closer to 3.  Our sample includes 14 clusters, which differ in inferred 
cooling-flow mass deposition rate by three orders of magnitude.  In 
every case, the isobaric cooling flow models overpredicts the observed 
spectrum at the lowest temperatures.  It appears that the {\it total} soft 
X-ray luminosity is roughly consistent with the predicted morphological 
mass deposition rate, but that the emission measure distribution is 
considerably steeper than this simple model would seem to require.

\section*{Empirical Parameterization}

We are able to fit the observed spectra reasonably well, however, using 
a variety of empirical parametrizations.  One that works especially well 
involves a simple modification of Equation (1):

\begin{equation}
\frac{dL}{dT} \approx \frac{5}{2} \frac{\dot{M} k}{\mu m_p} \frac{T}{T_a}
\end{equation}

\noindent
This means, for example, that at temperatures of $\frac{1}{4}$ of the 
ambient temperature of the cluster, the X-ray luminosity is weaker than 
predicted by the model by a factor of 4 or more.  The integrated soft 
X-ray luminosity is therefore consistent with that needed for massive 
cooling-flows to within a factor of 2, but the results are clearly 
discrepant with the standard multi-phase model at the lowest 
temperatures.  All clusters in the sample appear to follow this trend.

\section*{The General Theoretical Problem}

The cooling time for a constant pressure hot X-ray plasma is 
proportional to the temperature squared divided by the cooling 
function.  This implies that it would take 6 times longer for Abell 1835 
to cool from 8 keV to 3 keV than to cool from 3 to 0 keV.  It is 
extremely perplexing why there is no evidence for cooling after it has 
cooled 85$\%$ of the way.  The lack of emission lines at lower 
temperatures could imply that either there is no gas cooling or it could 
also mean that the cooling rate at lower temperature is much faster than 
predicted.  This could result from mixing processes which alter the rate 
of cooling at each temperature or from additional coolants which carry 
energy away in addition to the X-radiation.

The primary difficulty in finding a solution to this puzzle is that it 
is difficult to find a dynamical time-scale that would be so closely 
connected to the cooling time, which could vary by orders of magnitude 
depending on the local plasma conditions.  Various ideas have been 
suggested, but none appears to naturally explain the observed 
phenomena.  Mixing with cool clouds (\cite{jpeterson-B3:begelman}) 
condensed from the cooling-flow could liberate some energy at other 
wavelengths, posssibly in the UV, as suggested by H$\alpha$ observations 
(e.g. \cite{jpeterson-B3:heckman}, \cite{jpeterson-B3:crawford}) and O 
VI (\cite{jpeterson-B3:oegerle}), or through IR dust emission (e.g. 
\cite{jpeterson-B3:edge2}, \cite{jpeterson-B3:allen}).  However, one 
would also naively expect copious soft X-ray emission from the cloud 
interfaces with the hotter gas.  Some fraction of the thermal energy 
could be liberated in Alfv\'{e}n waves or mildly relativistic particles 
as the magnetic field compresses.  The very field lines that isolate the 
cooling blobs initially may reconnect and allow for modification of the 
cooling rate at each temperature.  Finally, some heating models (e.g. by 
AGN outflows \cite{jpeterson-B3:tabor}, \cite{jpeterson-B3:david}) could 
provide the necessary energy input to prevent the cooling, but they need 
to be finely tuned to work in just the right way for such a diverse 
sample of clusters.  

The solution to the soft X-ray problem could be complex and it could be 
difficult to distinguish between the alternatives.  Future work 
connecting the X-ray observations to other wavelength observations, 
further empirical quantification of the temperature distributions in 
cooling-flows, and considerably more theoretical progress may result in 
a more complete understanding of the cooling process.

\acknowledgements{
This work is based on observations obtained with XMM-Newton,
an ESA science mission with instruments and contributions directly funded by ESA Member States and the USA 
(NASA).}

\begin{figure*}[ht]
   \begin{center}
     \epsfig{file=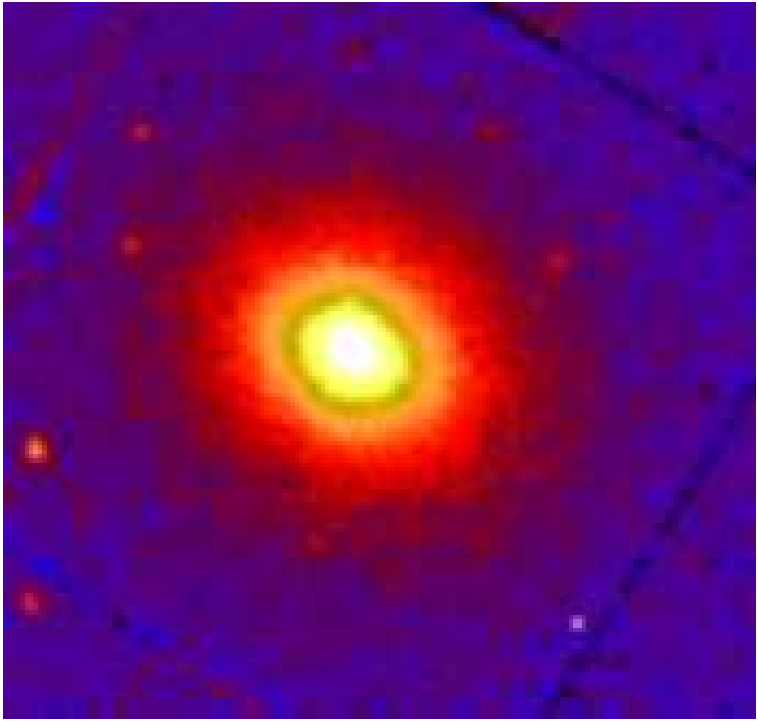, width=10cm}
   \end{center}
\caption{ An EPIC-MOS X-ray image for the cluster Abell S 1101. See 
Figure 2 for the corresponding RGS data.}
\label{jpeterson-B3_fig:fig1}
\end{figure*}
\begin{figure*}[ht]
   \begin{center}
     \epsfig{file=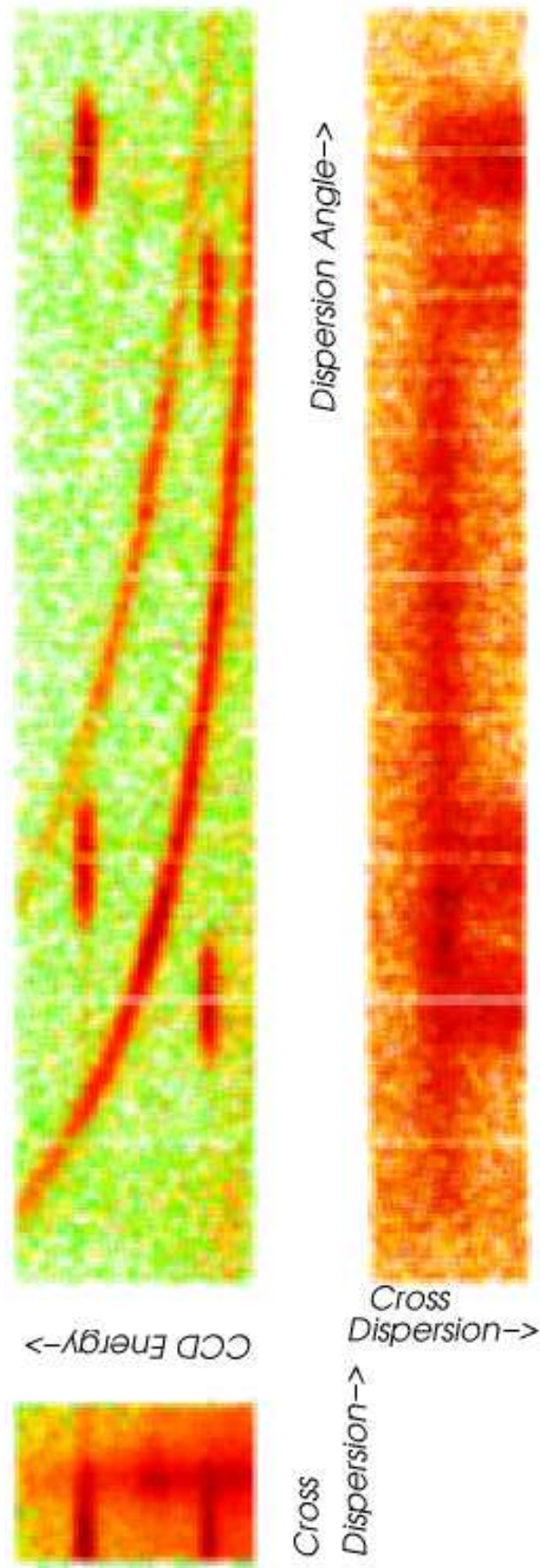, width=8cm}
   \end{center}
\caption{ The RGS data for Abell S 1101.  Shown is the three images for 
the three possible projections of the data.  The RGS data measures the 
dispersion coordinate, cross dispersion coordinate and CCD energy for 
each photon.  The dispersion coordinate vs. CCD energy histogram shows 
the first and second order dispersed spectrum (curved lines) and the 
four in-flight calibration sources.  The cross-dispersion vs. dispersion 
image shows the cluster dispersed along the 9 CCDs.  The data in the 
cross-dispersion direction corresponds to a one-dimensional dispersed 
image.}
\label{jpeterson-B3_fig:fig2}
\end{figure*}
\begin{figure*}[ht]
   \begin{center}
     \epsfig{file=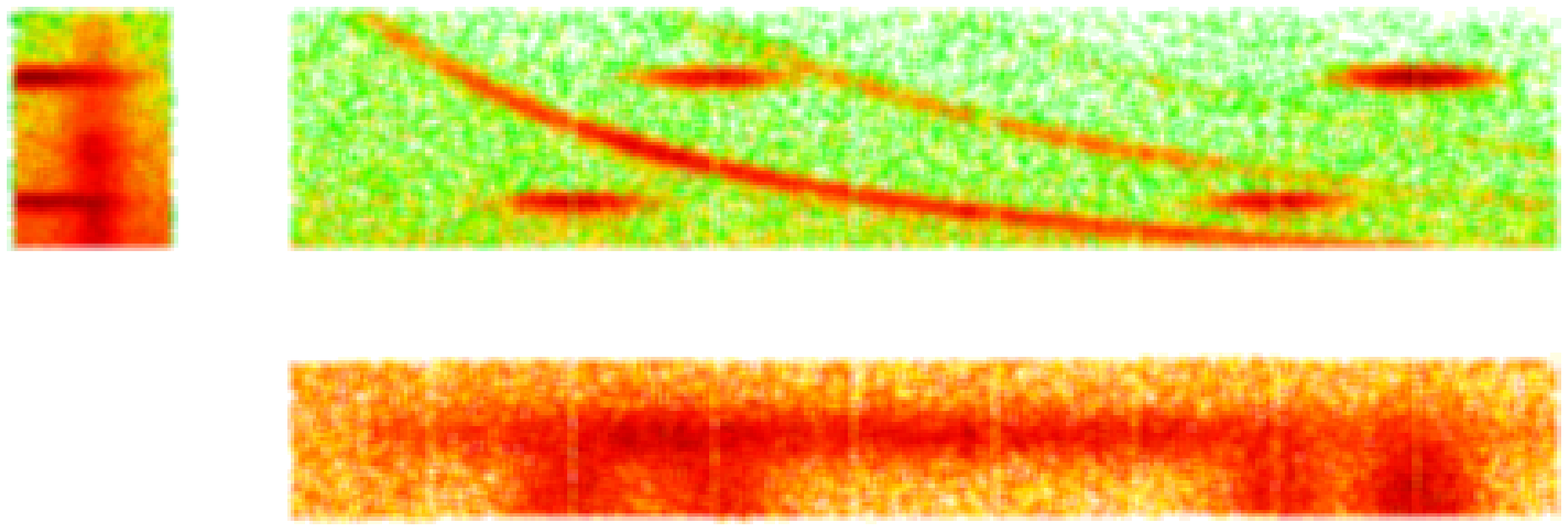, width=24cm,angle=90}
   \end{center}
\caption{ A simulation for Abell S 1101.  The images are the same as in 
Figure 2.  A Monte Carlo is used to generate the events and then they 
can be compared directly to the events in Figure 2.}
\label{jpeterson-B3_fig:fig3}
\end{figure*}
\begin{figure*}[ht]
   \begin{center}
     \epsfig{file=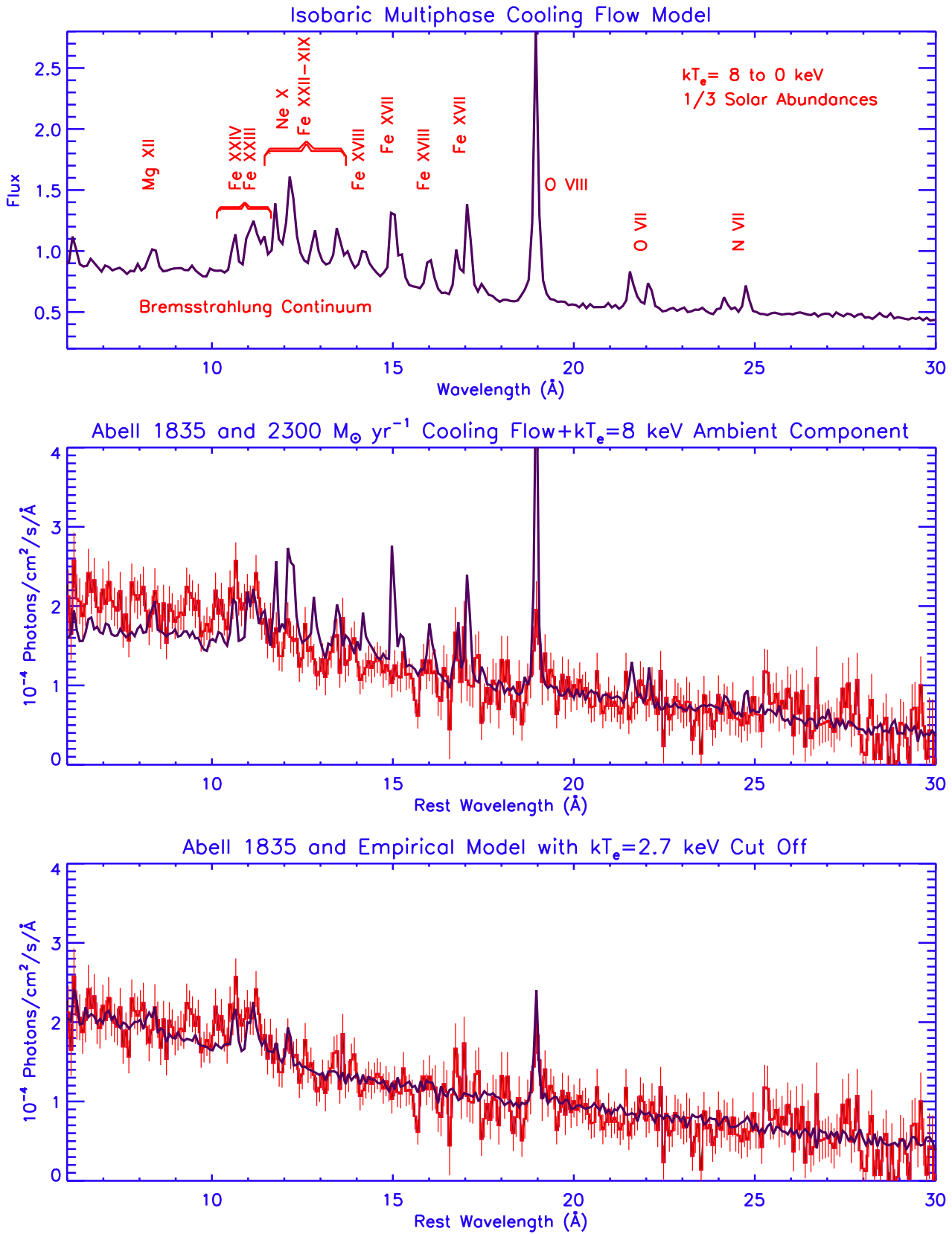, width=17cm}
   \end{center}
\caption{ The three panels show the prediction for the isobaric cooling 
flow model, the comparison of the model with the spectrum of Abell 1835, 
and a model where the temperature distribution is cut-off below 2.7 
keV.  A full explanation is in the text.}
\label{jpeterson-B3_fig:fig4}
\end{figure*}
\begin{figure*}[ht]
   \begin{center}
     \epsfig{file=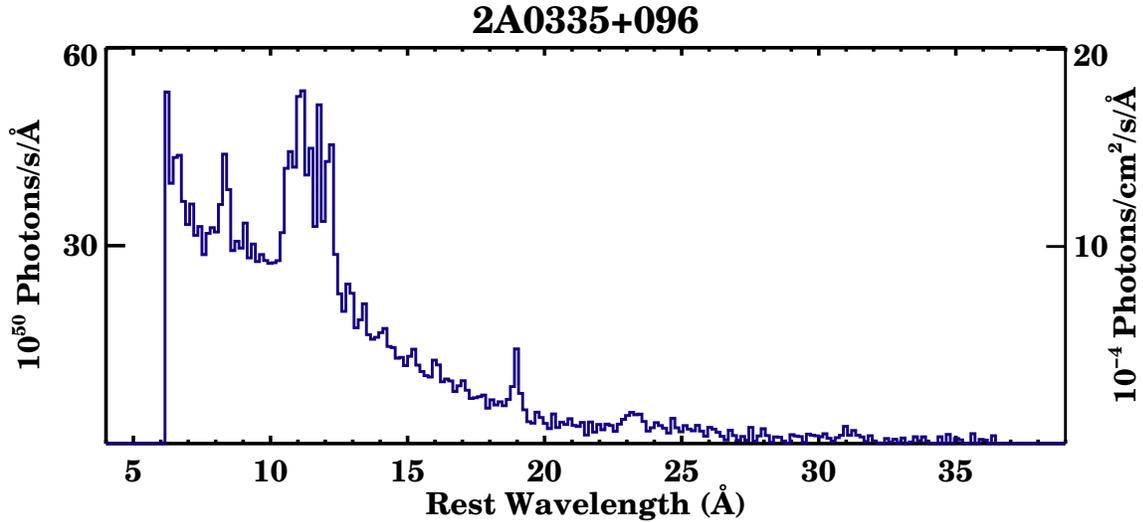, width=17cm}
   \end{center}
\caption{ The spectrum for the galaxy cluster 2A0335+096.  The spectrum 
is typical of intermediate temperature (3 keV) clusters.  The cluster 
lies close to the galactic plane so there is significant absorption of 
the spectrum.  Present in the spectrum is O VIII (19 \AA), Mg XII (8.4 
\AA), Si XIV \& XIII (6.2 \& 6.6 \AA), Fe XXIV-Fe XII (line blends 
between 10.6 and 12.2 \AA, Fe XXI-Fe XIX (blends near 12.8 and 13.6 
\AA), Fe XVIII (14.2 and 16.0  \AA), and Fe XVII (15 and 17 \AA).  The 
isobaric multi-phase model fails primarily because although low 
temperature ions such as Fe XVII are detected, they are much weaker than 
predicted.}
\label{jpeterson-B3_fig:fig5}
\end{figure*}
\begin{figure*}[ht]
   \begin{center}
     \epsfig{file=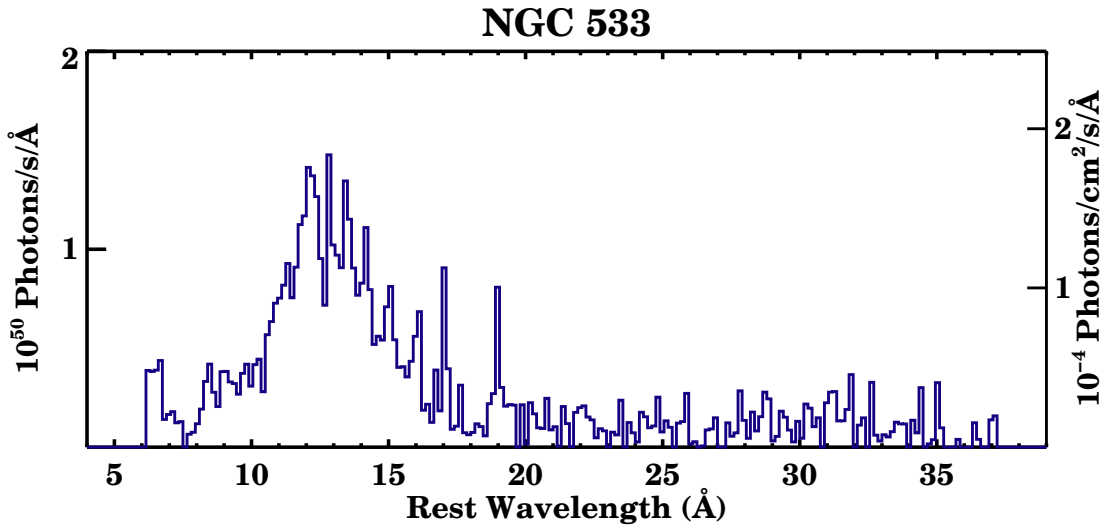, width=17cm}
   \end{center}
\caption{ The spectrum for the galaxy group, NGC 533.  The spectrum is 
qualitatively similar to other low temperature (1 keV) systems.  The 
line identifications are the same as the caption of figure 4.  The 
isobaric multi-phase model fails in these systems primarily because Fe 
XVII is weaker than predicted relative to Fe XVIII.}
\label{jpeterson-B3_fig:fig6}
\end{figure*}


\begin{thebibliography}{}

\bibitem[\protect
\astroncite{Allen et al.}{2001}]{jpeterson-B3:allen} Allen et al. 2001, 
MNRAS 322, 589.

\bibitem[\protect
\astroncite{Begelman \& Fabian}{1990}]{jpeterson-B3:begelman} Begelman, 
M. \& A. C. Fabian 1990, MNRAS 244, 26.

\bibitem[\protect
\astroncite{B\"ohringer et al.}{2001}]{jpeterson-B3:boehringer} 
B\"ohringer, H. et al. 2001, A \& A 365, 181.

\bibitem[\protect
\astroncite{Canizares et al.}{1979}]{jpeterson-B3:canizares1} Canizares, 
C. R. et al. 1979, ApJ 234, L33.

\bibitem[\protect
\astroncite{Canizares et al.}{1982}]{jpeterson-B3:canizares2} Canizares, 
C. R. et al. 1982, ApJ 262, L32.


\bibitem[\protect
\astroncite{Cowie \& Binney}{1977}]{jpeterson-B3:cowie} Cowie, L. L. \& 
J. Binney 1977, ApJ 215, 723.

\bibitem[\protect
\astroncite{Crawford et al.}{1999}]{jpeterson-B3:crawford} Crawford et 
al. 1999, MNRAS 306, 857.

\bibitem[\protect
\astroncite{David et al.}{2001}]{jpeterson-B3:david} David, L. P. et al. 
2001, ApJ 557, 546.

\bibitem[\protect
\astroncite{Edge et al.}{1999}]{jpeterson-B3:edge2} Edge et al. 1999, 
MNRAS 306, 599.

\bibitem[\protect
\astroncite{Edge}{2001}]{jpeterson-B3:edge} Edge, A. C. 2001, MNRAS 328, 
762.

\bibitem[\protect
\astroncite{Ettori et al.}{2002}]{jpeterson-B3:ettori} Ettori et al. 
2002, MNRAS, in press

\bibitem[\protect
\astroncite{Fabian \& Nulsen}{1977}]{jpeterson-B3:fabian2} Fabian, A. 
C. \& P. E. J. Nulsen 1977, MNRAS 180, 479.

\bibitem[\protect
\astroncite{Fabian}{1994}]{jpeterson-B3:fabian1} Fabian, A. C., ARA\&A 
1994, 32, 277.

\bibitem[\protect
\astroncite{Fabian et al.}{2001}]{jpeterson-B3:fabian3} Fabian, A. C. et 
al., MNRAS 320, 20.

\bibitem[\protect
\astroncite{Heckman et al.}{1989}]{jpeterson-B3:heckman} Heckman et al. 
1989, ApJ 338, 48.

\bibitem[\protect
\astroncite{den Herder et al.}{2001}]{jpeterson-B3:denherder} den Herder 
et al., A \& A 365, 7.

\bibitem[\protect
\astroncite{Johnstone et al.}{1992}]{jpeterson-B3:johnstone} Johnstone, 
R. M., MNRAS 255, 431-440.

\bibitem[\protect
\astroncite{Kaastra et al.}{2001}]{jpeterson-B3:kaastra} Kaastra, J. S. 
et al. 2001, A \& A 365, 99.

\bibitem[\protect
\astroncite{Kaastra et al.}{these proceedings}]{jpeterson-B3:kaastra2} 
Kaastra, J. S. et al., these proceedings

\bibitem[\protect
\astroncite{Molendi \& Pizzolato}{2001}]{jpeterson-B3:molendi} 
Molendi \& Pizzolato 2001, ApJ 560, 194.

\bibitem[\protect
\astroncite{Mushotzky \& Szymkowiak}{1988}]{jpeterson-B3:mushotzky} 
Mushotzky, R. F. \& Szymkowiak 1988, in Clusters of Galaxies, Vol. 299, 
53.

\bibitem[\protect
\astroncite{Oegerle et al.}{2001}]{jpeterson-B3:oegerle} Oegerle et al. 
2001, ApJ 560,  187.

\bibitem[\protect
\astroncite{Peterson et al.}{2001}]{jpeterson-B3:peterson1} Peterson, J. 
R. et al. 2001, A \& A 365, 104.

\bibitem[\protect
\astroncite{Peterson et al.}{in preparation}]{jpeterson-B3:peterson2} 
Peterson, J. R. et al., in preparation.

\bibitem[\protect
\astroncite{Peterson, Jernigan \& Kahn}{in preparation}]{jpeterson-B3:peterson3} Peterson, J. R., J. G. Jernigan, \& S. M. Kahn, in 
preparation.

\bibitem[\protect
\astroncite{Schmidt, Allen, \& Fabian}{2001}]{jpeterson-B3:schmidt} 
Schmidt, R. W., Allen, S. W., \& A. C. Fabian, MNRAS 327, 1057.

\bibitem[\protect
\astroncite{Tabor \& Binney}{1993}]{jpeterson-B3:tabor} Tabor, G. \& J. 
Binney 1993, MNRAS 263, 123.

\bibitem[\protect
\astroncite{Tamura et al.}{2001}]{jpeterson-B3:tamura} Tamura, T. et al. 
2001, A \& A 365, 87.

\bibitem[\protect
\astroncite{White et al.}{1991}]{jpeterson-B3:white} White, D. A. et al. 
1991, MNRAS 252, 1991.

\bibitem[\protect
\astroncite{Xu et al.}{2002}]{jpeterson-B3:xu} Xu, H. et al. 2002, 
astroph-0110013;
also Kahn, S. M. et al., these proceedings

\end{thebibliography}
\end{document}